\documentclass[]{aa}

\usepackage{natbib} 
\usepackage{txfonts}
\usepackage{graphicx}
\newcommand{\FBDA}{{\tt A}}
\newcommand{\FBEA}{{\tt B}}
\newcommand{\WABA}{{\tt C}}
\newcommand{\magunit}{mag arcsec$^{-2}$}
\newcommand{\densunit}{M$_{\sun}\ \mbox{pc}^{-2}$}

\newcommand{\smal}{half-major axis length}
\newcommand{\cun}{{\tt E}$_1$}
\newcommand{\cdeux}{{\tt E}$_2$}
\newcommand{\ctrois}{{\tt E}$_3$}
\newcommand{\cquatre}{{\tt F}$_1$} 
\newcommand{\ccinq}{{\tt F}$_2$}
\newcommand{\csix}{{\tt F}$_3$}
\newcommand{\rlun}{$L_{1,2}$}
\newcommand{\rlquatre}{$L_{4,5}$}

\newcommand{\rcor}{$R_\mathrm{CR}$}
\begin{document}

\title{The length of stellar bars in SB galaxies and N$-$body simulations}
\author{L\'eo Michel--Dansac\inst{1} \and\ Herv\'e Wozniak\inst{2}}
\offprints{L\'eo Michel--Dansac, \email{leo@astrosen.unam.mx}}
\institute{Instituto de Astronom\'{i}a, UNAM, Apartado Postal 877,
  22800 Ensenada, B.C., M\'exico \and Centre de Recherche Astronomique
  de Lyon, 9 avenue Charles Andr\'e, F-69561 Saint-Genis Laval cedex,
  France}
\date{Received/Accepted}
\authorrunning{Michel--Dansac \& Wozniak} 
\titlerunning{The length of stellar bars}
\abstract
{

\textit{Aims.\ } 
We have investigated the accuracy and reliability of six methods used to
  determine the length of stellar bars in galaxies or N-body simulations.

\textit{Methods.\ }
All these methods use ellipse fitting and Fourier decomposition of the
  surface brightness. We have applied them to N-body simulations that include
  stars, gas, star formation, and feedback. Stellar particles were
  photometrically calibrated to make B and K-band mock images. Dust absorption
  is also included.  We discuss the advantages and drawbacks of each method,
  the effects of projection and resolution, as well as the uncertainties
  introduced by the presence of dust.

\textit{Results.\ }
The use of N-body simulations allows us to compare the location of Ultra
  Harmonic Resonance (UHR or 4/1) and corotation (CR) with measured bar
  lengths. We show that the minimum of ellipticity located just outside the
  bulk of the bar is correlated with the corotation, whereas the location of
  the UHR can be approximated using the phase of the fitted ellipses or the
  phase of the $m=2$ Fourier development of the surface brightness.  We give
  evidence that the classification of slow/fast bars, based on the ratio
  ${\cal R} = R_{\mathrm{CR}}/R_{\mathrm{bar}}$ could increase from 1 (fast
  bar) to 1.4 (slow bar) just by a change of method.  We thus conclude that
  one has to select the right bar-length estimator depending on the
  application, since these various estimators do not define the same physical
  area.

}
\maketitle 
%--------------------------------------------------------------------
% INTRODUCTION
%--------------------------------------------------------------------
\section{Introduction}

Bars are ubiquitous and their importance for the long-term evolution of
galactic discs is well established nowadays (e.g. see reviews by
\citealt{SW93}, \citealt{KK04} and references therein). One of the
long-standing problems remaining for stellar bars is how to observationally
delimit the area of a bar, i.e. to find where a bar morphologically ends or,
in other words, how to measure its shape, length, and/or its width. Indeed,
determining many crucial observational quantities relies on the definition of
the bar size. However, a bar is not just an add-on morphological structure
embedded into an axisymmetric disc$+$bulge background. In fact, some
bulges (or pseudo-bulges, \citealt{KK04}) could be the result of the secular
evolution of bars, whereas part of the bar's stellar material spends a
significant fraction of time far out into the disc (the so-called 'hot'
population, \citealt{SS87}, \citealt{PF91}).  This is the main reason any
attempt to decompose 1D photometric profiles or 2D surface brightness in a
number of elementary contributions (bulge(s), disc(s), etc.) to recover the
'real' bar by substraction, in most cases, is doomed to failure.  

However, one of the simplest quantities used to estimate the strength of a bar
is its ellipticity.  For instance, \cite{M95} found a correlation between the
bar-axis ratio measured on blue photographic plates and star-formation
activity in a sample of 136 barred galaxies. \cite{Cetal99} confirm this
correlation using a more reliable photometry on red CCD images.  Using a
sub-sample of Martin's data, \cite{MtF97} emphasize that the most active
galaxies have the longer and thinner bars, but they also conclude that this is
only a necessary condition, not a sufficient one. However \cite{KPL02} argue
that this result could be dependent on the techniques used to determine bar
lengths and strengths. Moreover, the bar strength estimator $Q_b$ depends both
on the axis ratio and the mass of the bar \citep{ST80}.

There is another set of studies that needs accurate bar-length measurements.
Numerous authors have tried to correlate the size of morphological structures
to dynamical resonance locations. Circumnuclear and outer rings seem to be
correlated with the location of, respectively, the inner Lindblad resonance
(ILR) and the outer Lindblad resonance (OLR) (cf. \citealt{BC96}). The ratio
of the nuclear bar length to the large-scale bar length could be similar to
the ILR-to-corotation ratio, the nuclear bar corotation being dynamically
coupled with the large-scale bar ILR (e.g. \citealt{RS99}).  However, some
other simulations did not show such coupling (e.g. \citealt{hse01}). In short,
this matter is still being debated.

The extent of the bar (either nuclear or large-scale) should thus be compared
to the corotation location. There are several observational evidences
(\citealt{K90} and references therein) that stellar bars must end before the
corotation.  On the theoretical side, the theory of orbits \citep{C80},
hydrodynamic simulations (\citealt{ST80}, \citealt{A92}, \citealt{RT04},
etc.), and N-body simulations (\citealt{SS87}) all agree to predict that the
ratio of the corotation radius to the bar length is $1.2\pm 0.2$. Physically,
this limitation is due to an increase in the amount of chaotic orbits close to
the corotation.  Thus, near the corotation, there is no orbital support to
prolongate the shape of the bar.  However, it is difficult to observationally
confirm this value since it not only depends on the method used to determine
the pattern speed, but also on the criterion used to determine the bar length.
For instance, \citet{Aetal03} used four different criteria to determine the
bar length. Recent studies of the interaction between a stellar disc and a
live dark halo have widely discussed the evolution of the ratio of the
corotation radius over the bar length, implicitly outlining the importance
of an accurate bar-length determination (\citealt{DS00}, \citealt{AM02},
\citealt{A03}, \citealt{OD03}, \citealt{VK03}).

Another difficulty arises from the morphological difference between mass and
multi-wavelength surface-brightness distributions. Indeed, with N-body
simulations, the bar length is determined on the mass distribution, whereas
observational determinations use wavelength dependent surface-brightness
distributions. Comparisons between the two imply an understanding and an
estimation of the no-linear bias between mass and light. It poses, among
others, the question of the reliability of the various criteria used in the
literature and what they physically determine, depending on the support (mass
or light) on which they are applied.

To test the efficiency, reliability, and accuracy of various criteria for
determining the length of a bar, we decided to apply them to N-body
simulations that include stars, gas, and star-formation recipes. These
simulations were photometrically calibrated, taking dust extinction into
account (cf.  \citealt{MDW04}, hereafter Paper~I).  Thus, we could perform a
systematic comparison of methods applied to mass distributions, as well as to
photometrically calibrated mock images in blue and near-infrared wavelengths.
Then, the location of the bar ends, estimated using six different techniques,
was compared to the location of dynamical resonances. We thus exploit the
comprehensive knowledge of the dynamical properties of the simulation to find
the best bar-length estimator.

In Sect.~\ref{sec:criteria} we describe the six criteria used to measure bar
lengths. We also discuss some of their relative advantages and drawbacks.
Then, in Sect.~\ref{sec:model}, we describe our numerical models and their
evolution. We also briefly recall the technique of photometric calibration
used in Paper~I. The six criteria are applied to our simulations in
Sect.~\ref{sec:tests}.  We also analyse inclination effects on our results in
this section.  Sect.~\ref{sec:results} deals with the temporal evolution of
bar lengths.  Correlations between the various bar length determinations and
the location of dynamical resonances are studied in
Sect.~\ref{sec:discussion}. The distinction between fast and slow bars and the
limitations due to the lack of dark halo in our simulations are discussed in
the same section. We finally summarise our results in the last section.

%--------------------------------------------------------------------
\section{Measurements of bar length}
\label{sec:criteria}

The method adopted by \citet{M95} for measuring the length and width
of bars is exclusively visual and relies on photographic prints in the
blue band. He estimates that the uncertainty is about 20\% with this
method. He defines the half-major axis of the bar as the length 
from the galaxy centre to the sharp outer tip where spiral arms begin,
and the half-minor axis as the length from the centre to the edge of
the bar, oval, lens, or spheroidal component, measured perpendicularly
to the major axis. Using CCD red images, \citet{Cetal99} made a more
sophisticated numerical analysis. They first extracted a 3-pixel wide
photometric profile along the major axis of the bar. The half-major
axis of the bar is the distance from the centre of the galaxy to
where the bar obviously ends. This is where the surface-brightness
profile changes slope abruptly to become steeper; this also coincides
with the origin of the spiral arms. This measurement is not automatic,
as it relies on a subjective judgement of where the bar ends, and is
comparable to that of Martin (1995).

Beyond eye estimates, several attempts to find an automatic and objective
criterion were made to determine where the bar ends and the disc or the
spiral arms start.  The objective criteria can be classified in two groups
defined by the tools that used: 3 based on ellipse fitting (noted
\cun\ to \ctrois), and 3 based on the Fourier analysis (noted \cquatre\ to
\csix).  We applied all six criteria on the mass distribution and on the
calibrated images.

From now on, we use the expression `bar radius' instead of `bar length' since
the various criteria used in this paper deal with either half-major axis
length or radius. Each measurement must obviously be doubled whenever
the real `length' or the `diameter' should be determined.

\subsection{Ellipses fitting}
\label{ssec:ellipse}

In the past, the analysis of isophotal shapes using a functional form was 
done using either classical ellipses (e.g. \citealt{WP91}, \citealt{Wetal95})
or generalised ellipses \citep{Aetal90}, which permits the description of
boxy-like isophotes. 

By fitting ellipses to the isophotes of a large sample of barred galaxies,
\citet{Wetal95} determined some general rules about the radial behaviour of
both the ellipticity (defined by $e\!=\!1-b/a$ where $a$ and $b$ are the
half-major and half-minor axis lengths respectively) and the position-angle
(PA) of the isophotes. The ellipticity is minimum or vanishes at the centre,
because of either seeing effects or a luminous spherical bulge; then it
increases to reach a maximum $e^\mathrm{max}$, often at about the middle of
the bar, and then progressively decreases towards $e^\mathrm{min}$ at the
place where the isophotes should become axisymmetric (disc) in the face-on
case.  Of course, $e^\mathrm{min}$ is determined by the galaxy inclination
and/or the real non-axisymmetric shape of the disc.  The PA is constant along
the bar and, in general, sharply takes another constant value for the disc
isophotes.  If a spherical bulge is present, $e$ will only start to increase
outside the bulge.

Three criteria can be defined using the results of ellipse fitting. For the
first criterion (\cun), the bar radius is the radius where the ellipticity
profile reaches a maximum. It has been introduced by \citet{WP91}, who studied
a sample of SB0 galaxies in optical wavebands. Although this criterion is very
useful for automated measurements (e.g. \citealt{RE97}, \citealt{J97},
\citealt{L02}) because the maximum ellipticity is always clearly defined in
SBO galaxies, it seems to give rather short values of bar radius when it is
compared to eye estimates (cf. also Fig.~\ref{fig:profiles}).

\citet{Wetal95} later introduced another criterion (hereafter called \ctrois)
given by the position where the PA variation exceeds 5\degr\ {\it and}
ellipticity drops to disc values (rounder isophotes). This criterion gives
higher values that were expected to be upper limits of real bar radii.
\citet{ES03} used a PA variation of 10\degr\ instead of 5, which should lead
to slightly longer bars. The regions where the \ctrois\ criterion is satisfied
are generally less dusty than for \cun\ since gas density strongly decreases
outside the bar. The bar-radius estimation using \ctrois\ might be more robust
for most galaxies than \cun\ with respect to extinction but, depending on the
detailed spatial distribution of the gas, could also lead to large errors.  A
variant of the \ctrois\ criterion has been used by \citet{Jetal04} to analyse
almost 1\,500 barred galaxies in the ACS GEMS survey, showing its reliability
in data-mining studies.

\begin{figure}
\resizebox{\hsize}{!}{\includegraphics{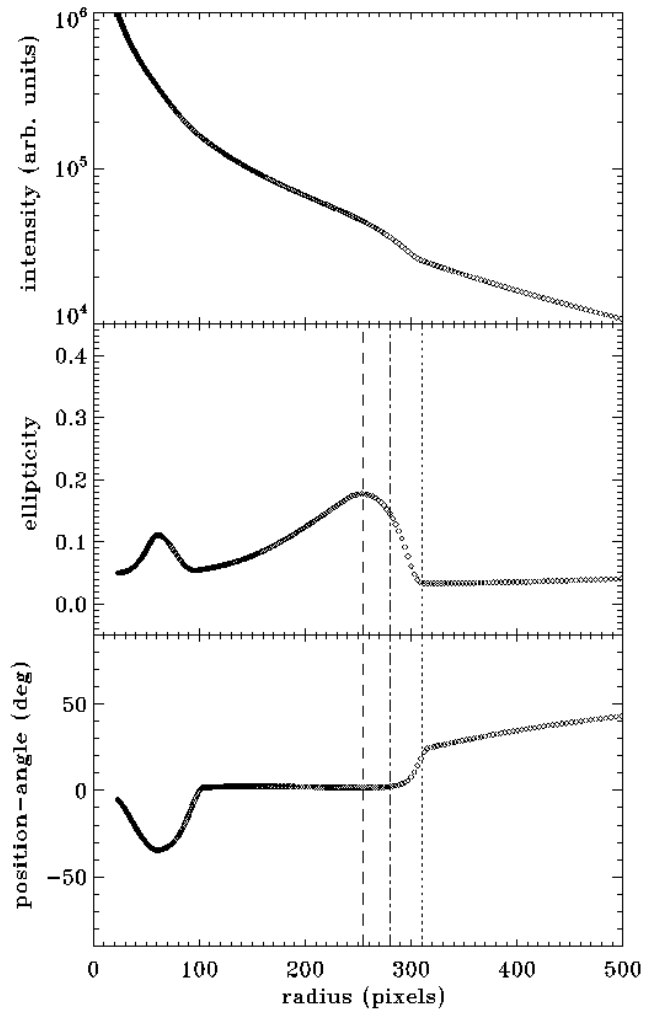}}
\caption{Illustrative example of the profiles obtained by ellipse
fitting. From top to bottom: surface brightness in arbitrary units,
ellipticity, and PA (in degrees) profiles as a function of
half-major axis of the fitted ellipse. Each point represents a fitted
ellipse. The vertical lines represent bar radii determined with \cun\
(dashed line), \cdeux\ (dot-dashed line), and \ctrois\ (dotted line).}
\label{fig:ellfit}
\end{figure}

We now define a new criterion (\cdeux) that is also based on ellipse fitting.
In the region of the bar, where the ellipticity profile looks like a bump, the
PA is approximately constant, hence the PA profile forms a plateau
(cf.~Fig.~\ref{fig:ellfit}).  For the \cdeux\ criterion, bar radius is
measured at the end of this plateau. This corresponds to a twist between
isophotes of the bar and isophotes of the disc region where spiral arms begin.
This criterion gives an estimation of bar radii between those of \cun\ and
\ctrois\ criteria in the absence of dust extinction.

To illustrate how these criteria could lead to different bar-radius
estimations, we made an artificial image that contains a bulge,
two embedded bars, and a disc. Our purpose is not to make an extensive
study, varying each parameter to quantify the quality of each bar-radius
estimator. 
The bulge has a de Vaucouleurs profile of central intensity
$I_e=9000$, scale length $R_e=200$. It is slightly elongated along
PA~$=0\degr$ with an ellipticity of 0.05. The exponential disc has a
central intensity $I_d=8000$, scale length $R_d=300$, and an
ellipticity of 0.1. The PA of the disc's main axis is
PA~$=60\degr$. The two bars have a surface density:
\begin{displaymath}
I(x,y)=I_i \left( 1- \left|\frac{x}{a_i}\right|^{c_i} -
\left|\frac{y}{b_i}\right|^{c_i} \right)
\end{displaymath}
where $a_i$ is the half-major axis (i.e. bar radius), $b_i$ the half-minor
axis, $I_i$ the central intensity, and $c_i$ the shape parameter
\citep{Aetal90}. The main bar has $a_1=300$, $b_1=200$, $I_1=5000$, and
$c_1$ varies from 2 at the centre (perfect ellipses) to 3.5 at the ends,
thus giving a rectangular-like bar. It is aligned with the $x$-axis. A
nuclear bar is added at PA~$=-45\degr$ with the same intensity profile
with parameters $a_2=100$, $b_2=50$, $c_2=2$ everywhere, and $I_2=16000$.
All scale units are pixels and intensity units are numbers that could
be considered as ADU.

We display the results of the ellipse fitting in Fig.~\ref{fig:ellfit}, i.e.
the ellipticity and PA of the isophotes as a function of the half-major axis
length of the fitted ellipse. Applying the above criteria leads to a bar
radius of 255 for \cun, 280 for \cdeux, and 310 for \ctrois. This last
criterion slightly overestimates the real bar radius because of the
rectangular shape of the isophotes near the end of the bar. Ellipses are
indeed not suited to strong rectangular bars.

\subsection{Fourier analysis}
\label{ssec:fourier}

Fourier analysis of the surface brightness \citep{o90} is based on
the decomposition of azimuthal density profiles, $I(r,\theta)$, into a Fourier
series
\begin{displaymath}
I(r,\theta) = I_{0}(r) + \sum_{m=1}^{\infty}
\left[A_{m}(r)~\cos(m\theta) + B_{m}(r)~\sin(m\theta)\right]
\end{displaymath}
where, for $m\neq 0$, the coefficients are given by:
\begin{displaymath}
A_{m} = \frac{1}{\pi}~\int_{0}^{2\pi}
I(r,\theta)~\cos(m\theta)~d\theta
\end{displaymath}
\begin{displaymath}
B_{m} = \frac{1}{\pi}~\int_{0}^{2\pi}
I(r,\theta)~\sin(m\theta)~d\theta.
\end{displaymath}
The Fourier amplitude of the $m$th component ($m>0$) is defined as:
\begin{displaymath}
I_{m}(r) = \sqrt{A_{m}^{2}(r)+B_{m}^{2}(r)}.
\end{displaymath}
The bar--interbar contrast is then computed as:
\begin{equation}
\label{eq:contrast}
{\cal C} =I_\mathrm{b}/I_\mathrm{ib} = (I_0 + I_2 + I_4 + I_6) / (I_0 - I_2
+ I_4 - I_6).
\end{equation}

\begin{figure}
\resizebox{\hsize}{!}{\includegraphics{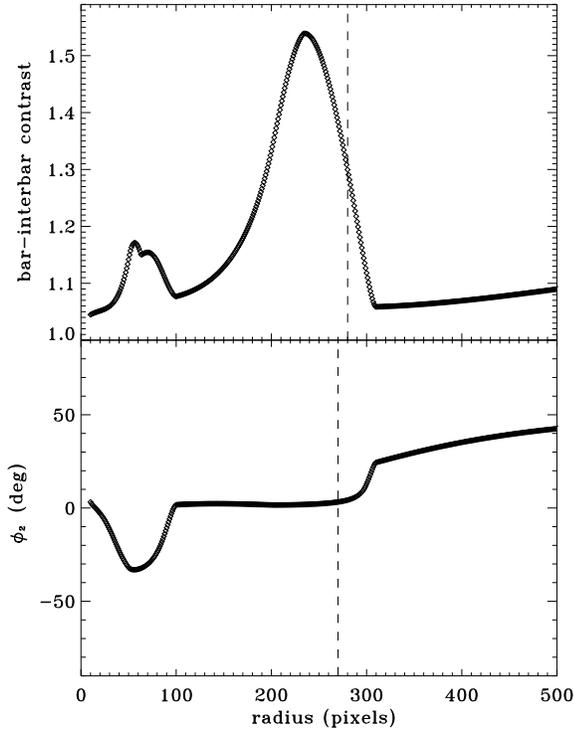}}
\caption{Illustrative example of the profiles obtained by 
Fourier analysis: bar--interbar contrast (top) and phase of the $m=2$ component
(bottom) as a function of radius. Each point represents a pixel. The dashed
vertical lines symbolise the end of the bar determined using \ccinq\ (top) and
\csix\ (bottom) criteria.}
\label{fig:foufit}
\end{figure}

We display ${\cal C}$ in Fig.~\ref{fig:foufit} for the test image used
above. Typically, $\cal C$ increases steeply and reaches its peak
value in the bar region, then falls toward the bar end.  The bar
region is defined by \citet{o90} as the zone where the contrast
exceeds 2 (\cquatre\ criterion hereafter). This criterion has been
revisited by \citet{Aetal00}, who redefine the end of the bar region
where the contrast reaches the full width at half maximum (criterion
\ccinq). In other words, the bar ends where
${\cal C} = 0.5 \left[ \max({\cal C}) + \min({\cal C}) \right]$. 

Another criterion (\csix) is determined using the radial profile of the phase
of the $m=2$ component, which is defined as:
\begin{displaymath}
\phi_{2}(r) = \arctan[B_{2}(r)/A_{2}(r)].
\end{displaymath}
We display this phase profile in Fig.~\ref{fig:foufit} for the test image.
The $m=2$ phase can be interpreted as the PA of the features that dominate
the $m=2$ component. Thus, the phase successively gives the PA of the bar,
then of the two-arm spiral structure, if present, or/and the PA of the disc (if
not perfectly disky, which is always the case, due to inclination at
least). The $m=2$ phase profile is then similar to the PA profile, and the
phase in the bar region is approximatively constant. We measure bar radius at
the end of this plateau.  This criterion has already been applied to mass
distribution from N-body simulations (e.g. \citealt{DS00}), as well as to
observations (e.g. \citealt{Aetal03}).

These criteria applied to the artificial image lead to a bar radius of $\approx
280$ for \ccinq, whereas \csix\ gives $\approx 270$. However, \cquatre\ fails
to give any estimation, since the contrast never exceeds the value of 2. Thus,
for the case of our test image, \ccinq\ and \csix\ underestimate the radius of
the main bar.

%-------------------------------------------------------------------
% DESCRIPTION DES SIMULATIONS
%-------------------------------------------------------------------
\section{Numerical models}
\label{sec:model}

\subsection{hydro$+$N-body simulations}

We used {\tt PMSPHSF}, the N-body code developed in Geneva. It includes stars,
gas, and recipes to simulate star formation.  The broad outlines of the code
are as follows: the gravitational forces are computed with a particle--mesh
method using a 3D polar grid \citep{PF93}, the hydrodynamics equations are
solved using the smooth particle hydrodynamics (SPH) technique (see
\citealt{FB93} for this implementation and \citealt{Mo92} for a review of the
method), and the star-formation process is based on Toomre's criterion for the
radial instability of gaseous discs \citep{FB95}. For the present work, the
radiative cooling of the gas was computed assuming a solar metallicity
for Runs \FBDA\ and \FBEA\ and a cosmological metallicity for Run \WABA.  The
metallicity change of gas particles was computed assuming net yields given by
\citet{M92}. At birth, star particles are created from gas, so they have the
same metallicity.

An initial stellar population was set up to reproduce a disc galaxy with an
already formed bulge. These particles form what we call hereafter the `initial
population', as opposed to particles created during the evolution (`new
population').  The initial stellar positions and velocities for $N_s$
particles of the same mass are drawn from a superposition of two axisymmetric
Miyamoto-Nagai discs of mass $M_1$ and $M_2$, of scale lengths
$a_{1}+b$ and $a_{2}+b$, respectively, and identical scale height $b$ (cf.
Table~\ref{tab:param}, where masses are in M$_{\sun}$ and lengths in kpc).
Their gravitational potential is then given by \citep{mn75}:
\begin{displaymath}
\Phi_{1,2}(R,z) = -
\frac{GM_{1,2}}{\sqrt{R^{2}+\left(a_{1,2}+\sqrt{z^{2}+b^{2}}\ \right)^{2}}}
\end{displaymath}
where $R=\sqrt{x^2+y^2}$.  The gaseous component is represented by $N_g$
particles for a total mass $M_g$, scale length $a_g+b_g$, and scale height
$b_{g}$.  As in Paper~I, we did not include a dark halo, to be able in the near
future to compare the effects of having added such an additional component.

\begin{table*}
\centering
\begin{minipage}{126mm}
\caption{Initial parameters of the simulations}
\label{tab:param}
\begin{tabular}{@{}llllllllllll@{}}
\hline \hline
Run & $N_s$           & $N_g$     & $M_1$      & $M_2$   & $M_g$         &$a_1$ & $a_2$ &$b$ & $a_g$ & $b_{g}$ \cr
      &               &           &            &         &               &       &       &      &       &          \cr
\hline
\FBDA & 500\,000      & 50\,000   &10$^{10}$   &10$^{11}$&1.1\,10$^{10}$ & 0.5   & 3.0   & 0.5  & 3.0   & $10^{-4}$\cr
\FBEA & 500\,000      & 10\,000   &10$^{10}$   &10$^{11}$&3.66\,10$^{10}$& 0.5   & 3.0   & 0.5  & 3.0   & 0.5      \cr
\WABA & 600\,000      & 10\,000   &2\,10$^{10}$&10$^{11}$&4\,10$^{10}$   & 0.5   & 5.0   & 1.0  & 10.0  & 0.5      \cr
\hline
\end{tabular}
\end{minipage}
\end{table*}

\begin{figure}
\resizebox{\hsize}{!}{\includegraphics{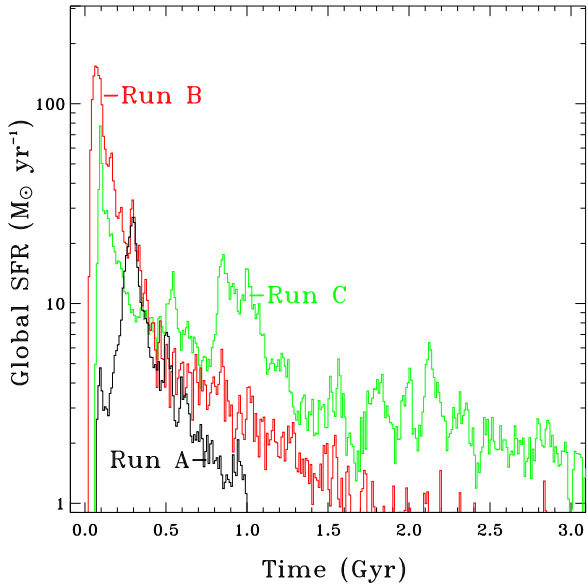}}
\caption{Evolution of the star-formation rate (SFR) of Runs \FBDA\
(black), \FBEA\ (red), and \WABA\ (green).}
\label{fig:sfr}
\end{figure}

\begin{figure*}
\centering
\includegraphics[height=17cm]{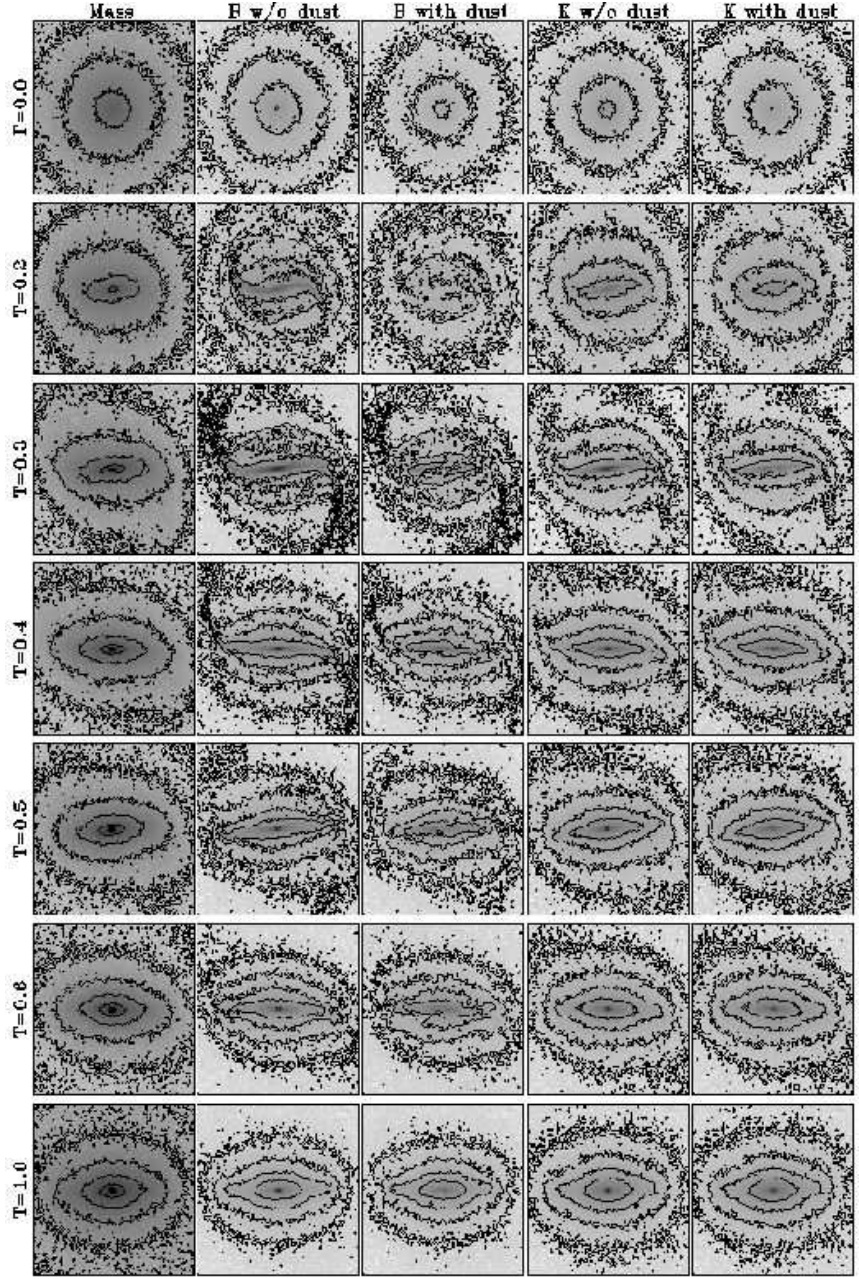}
\caption{Evolution of Run \FBDA\ from $t=0$ to $t=1$~Gyr. From left to
right are displayed the mass distribution (in $\log$ of \densunit) and
the calibrated images (B dust-free, B with extinction, K dust-free, and
K with extinction) in \magunit. The field of view (10~kpc) is the same
for each frame. The particles have been rotated so that the bar
position is roughly horizontal. Greyscale images and isocontours of
mass surface density range from $10^{1}$ to $10^{5.2}$~\densunit. Mass
isocontours are logarithmically spaced by 0.47~$\log$(\densunit). The B
greyscale images range from 11 to 26 \magunit, while contours range
from 20 to 23 \magunit\ and are spaced by 1 mag. The K images range from 9
to 22 \magunit; K isocontours range from 17 to 20 \magunit\ and are
spaced by 1 mag.}
\label{fig:contour}
\end{figure*}

Three simulations were used. Run \FBDA\ (cf. Fig.~\ref{fig:contour}) was
computed over 1 Gyr. It was analysed in some detail in Paper~I. Runs
\FBEA\ (Fig.~\ref{fig:contour1}) and \WABA\ (Fig.~\ref{fig:contour2}) were
performed on a longer timescale. They will be used to check our results
on the long term with different initial conditions. The main differences
between the three simulations are the initial mass of gas and its spatial
distribution.  The resulting star-formation histories are quite different (cf.
Fig.~\ref{fig:sfr}): Run \FBDA\ is more representative of a short timescale
burst of star formation; Run \WABA\ shows several small events but star
formation is sustained over a longer period because of the more massive
gaseous reservoir initially distributed over a greater length scale than for
Run \FBDA; Run \FBEA\ is an intermediate case.

During the three simulations, a bar progressively grows and then evolves
secularly. The gravitational torques due to the bar and spiral structures
drive the gas inwards. Star formation occurs along the transient spiral
structure, along and around the bar, and in the central parts. The bar
formation is thus associated to the starburst seen in Fig.~\ref{fig:sfr}. The
bar-formation timescale differs from one simulation to the next, as a
consequence of the different initial conditions. Another consequence is the
morphological difference in the resulting bars. With these simulations, we
have thus obtained a collection or 'sample' of snapshots with a wide variety
of bar shape, as can be seen in
Figs.~\ref{fig:contour},~\ref{fig:contour1}, and \ref{fig:contour2}; note the
larger field of view used in Fig.~\ref{fig:contour2} than in the other two.
Moreover, our simulated bars have properties (such as ellipticity or strength)
representative of nearby galaxies, as discussed in Sect.~\ref{sec:tests}.

\begin{figure*}
\centering
\caption{Like Fig.~\ref{fig:contour} but for Run \FBEA\ from $t=0$ to
$t=5$~Gyr.  Greyscale images and isocontours of mass surface density
range from $10^{1}$ to $10^{5.2}$~\densunit. Mass isocontours are
logarithmically spaced by 0.47~$\log$(\densunit). The B greyscale images
range from 12 to 27 \magunit, while contours range from 19 to 23
\magunit\ and are spaced by 1 mag. The K images range from 10 to 23
\magunit; K isocontours range from 16 to 20 \magunit\ and are spaced
by 1 mag.}
\label{fig:contour1}
\end{figure*}

\begin{figure*}
\centering
\caption{Like Fig.~\ref{fig:contour} but for Run \WABA\ from $t=0$ to
$t=5$~Gyr.  The field of view is 16~kpc. Greyscale images and
isocontours of mass surface density range from $10^{1}$ to
$10^{5}$~\densunit. Mass isocontours are logarithmically spaced by
0.44~$\log$(\densunit). The B greyscale images range from 13 to 27
\magunit, while contours range from 19 to 23 \magunit\ and are spaced
by 1 mag. The K images range from 10 to 23 \magunit; K isocontours range
from 16 to 20 \magunit\ and are spaced by 1 mag.}
\label{fig:contour2}
\end{figure*}

\subsection{Mock calibrated images}

Dynamical properties of a bar obviously depend on the mass distribution.
Collisionless N-body simulations are thus one of the best tools for studying
them. However, when one deals with determining observational properties
(i.e. photometric properties in our case), an additional dependency makes the
use of collisionless N-body simulations much more difficult: the mass-to-light
ratio (M/L).  Indeed, M/L varies in time and space since it depends both on
the star-formation history and on the mixing of stellar populations by secular
evolution. Simulations with gas and star formation permit photometrically
calibrated light distributions to be obtained by performing a post-processing
calibration with stellar-population synthesis models, as described hereafter. 

For each stellar particle, given its age and metallicity, the mass-to-light
ratio was obtained from a bi-linear interpolation of the tables of GISSEL2000
\citep{BC93}, for a Salpeter initial mass function with a lower mass cutoff
$M_d = 0.1$~M$_\odot$ and a upper cutoff $M_u = 100$~M$_\odot$.

For the initial population of Run \FBDA\ (present at the beginning of the
simulation), we assumed an age of 10.4~Gyr and metallicity of $Z=0.004$,
like the model A of Paper~I. In that paper, we considered various
calibrations of the initial population and found that the length of the bar
(determined with the \ctrois\ criterion, defined above) is almost independent
of the calibration model. We thus decided here to restrict our study to one
calibration model for the sake of clarity.

Obviously, the assumptions on the age and metallicity of the initial
population implicitly imply that all particles were simultaneously born
10.4~Gyr before the beginning of the simulation.  Assuming an observer located
at $z=0$ at the end of the simulation ($t=1$~Gyr), this calibration implies a
redshift of formation of $z=3$.  For the longer Runs \FBEA\ and \WABA, we
considered several sets of initial population ages and metallicities. For
clarity, we deal only with the calibration model that assumes an initial
population age of 4.4~Gyr and a metallicity of $Z=10^{-4}$. This leads to a
final age of 11.4~Gyr, the same as for Run \FBDA.

In Paper~I we worked with four wavebands to cover the wavelength domain from
the visible to the near infrared, namely B, R, H, and K Johnson. Here, we
restrict our analysis to B and K bands without any loss in generality.

Figures \ref{fig:contour} to \ref{fig:contour2} display the evolution of the
mass density and the B and K calibrated images with and without dust, in the
bar region for a selection of snapshots. For Run \FBDA, large scale snapshots
of the evolution can be found in Paper~I. These figures clearly show that the
disc undergoes a complete redistribution of the mass.  For instance, for Run
\FBDA\, after 1~Gyr, the total mass inside the central kpc has roughly
doubled.  The first cause of this mass inflow is the formation of a stellar
bar. Then, due to the gravitational torques exerted on the gas by the stellar
bar, the extra mass in the form of gas and new stars amounts to
$3.5\,10^{9}$~M$_{\sun}$ at $t=1$~Gyr for Run \FBDA, which is only 30\% of the
whole additional mass. In fact, the initial stellar population contributes to
the other 70\%.

After individual particle photometric calibration, mock CCD images were
obtained summing particle luminosities into a 512$\times$512 pixels grid. The
field of view was 60~kpc, which gave a spatial resolution of $\approx 117$~pc,
1.3 times our smallest $N$-body grid resolution. We thus produced one frame
per waveband and per snapshot of the simulation. Our results are obviously
independent of the bar PA with respect to the North or any other axis. We thus
decided to systematically rotate the positions of particles to align the bar
with the x-axis.

To mimic real observations we should have to convolve our images with a
point-spread function. However, this last stage depends on the telescope and
observation-site characteristics. It thus introduces a few free parameters
that cannot be constrained without any detailed comparisons with real
observations, which is not our purpose.

\subsection{Dust extinction}

Dust extinction in B and K bands was simulated assuming a constant gas-to-dust
ratio
\begin{displaymath} 
N(\ion{H}{i})/A_\mathrm{V} = 5.34\times 10^{21}\mbox{cm}^{-2}.
\end{displaymath}
Then, $A_\mathrm{V}$ is converted to $A_{B}$ and $A_{K}$.  Extinction
was computed in a cube with the same spatial resolution as for our
images and 11 slabs along the line-of-sight (cf. Paper~I for more
details). Each slab absorbs the stellar luminosity behind it.  For
each slab, the gas density distribution is obtained by convolving particle
positions with the SPH kernel.

%-------------------------------------------------------------------
% COMPORTEMENT DES CRITERES SUR LES SIMULATIONS
%-------------------------------------------------------------------
\section{Tests of bar-radii criteria on numerical simulations}
\label{sec:tests}

\begin{figure*}
\centering
\resizebox{\hsize}{!}{\includegraphics{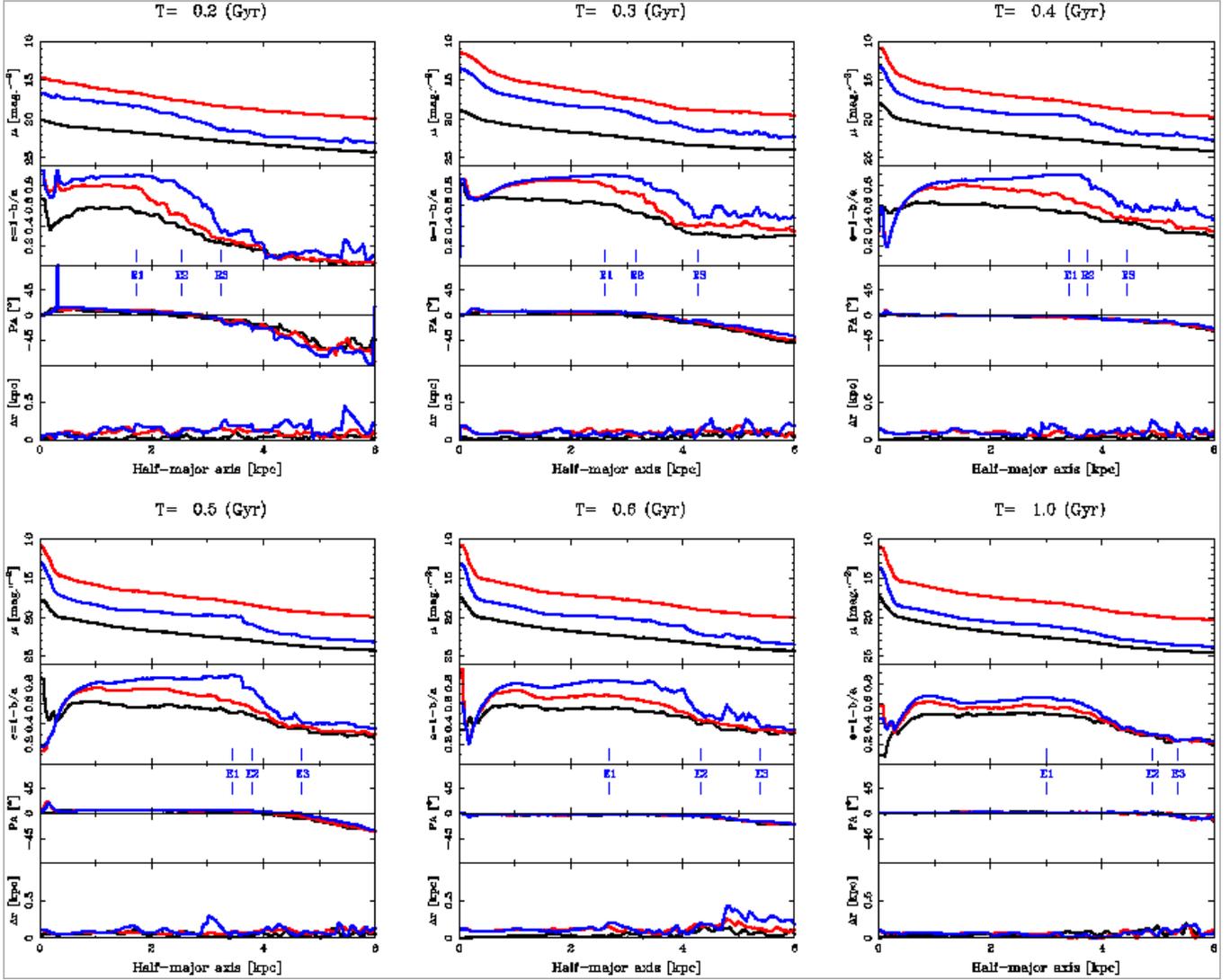}}
\caption{In each panel, for the two bands (B plotted in blue and K in red) and
the mass distribution (in black) for Run \FBDA: from top to bottom, surface
brightness ($\mu$), ellipticity (e=1-b/a), PA, and central
shift for each fitted ellipse. The range in radius
for the figures begins at 1 px, i.e. 117~pc; the upper bound
corresponds to a region that encloses the bar during the entire 
run. The mass density profiles have been scaled so that the 30$^{th}$
magnitude corresponds to 1~\densunit. The labels \cun, \cdeux, and
\ctrois\ refer to the criteria used to determine the bar radius in
B band. See the text for their definitions.}
\label{fig:profiles}
\end{figure*}

\begin{figure*}
\centering
\resizebox{\hsize}{!}{\includegraphics{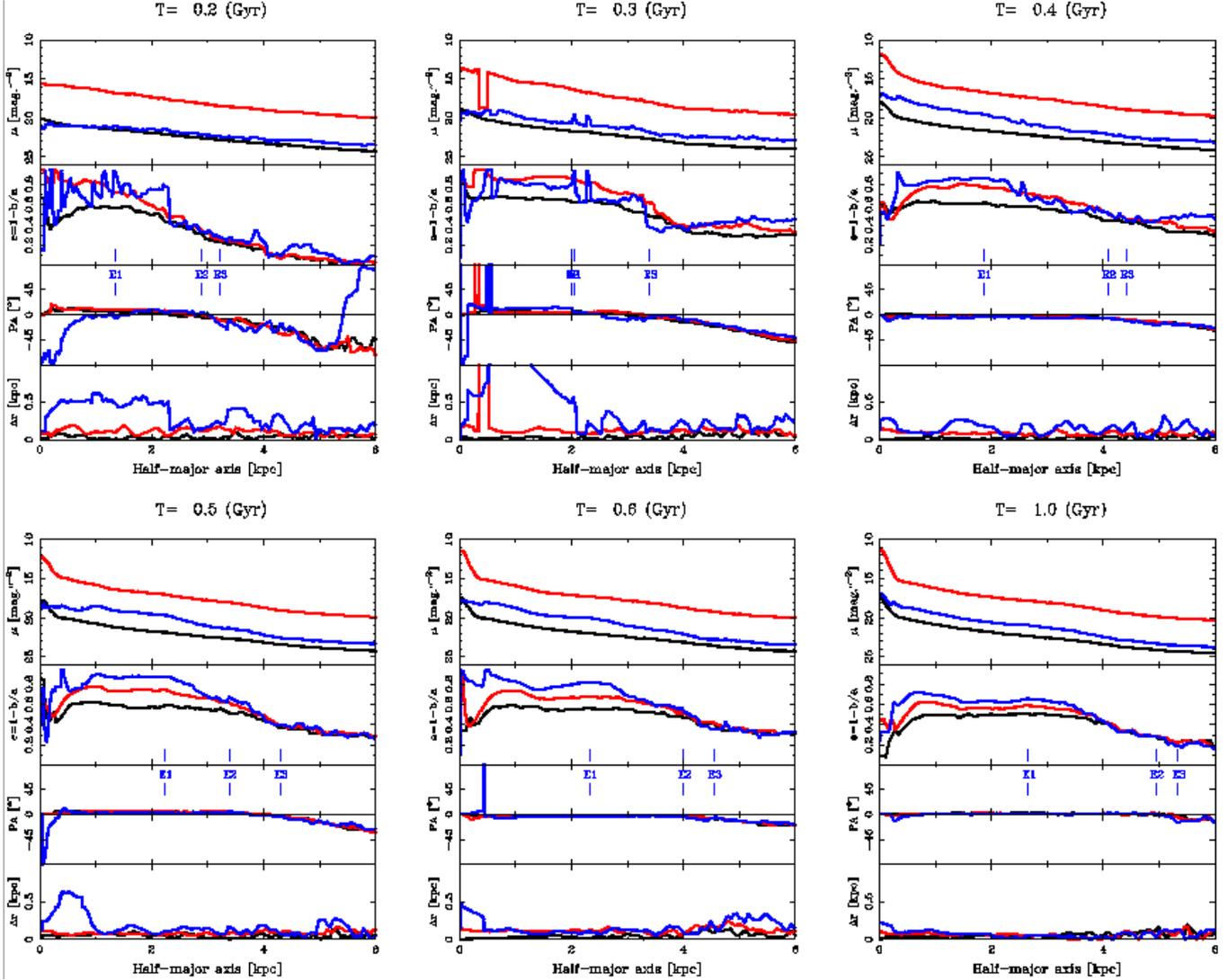}}
\caption{Like Fig.~\ref{fig:profiles} but for dusty images. Results for
the mass density are the same than in Fig.~\ref{fig:profiles} since
they are not affected by extinction.}
\label{fig:profiles2}
\end{figure*}

\begin{figure*}
\centering
\resizebox{\hsize}{!}{\includegraphics{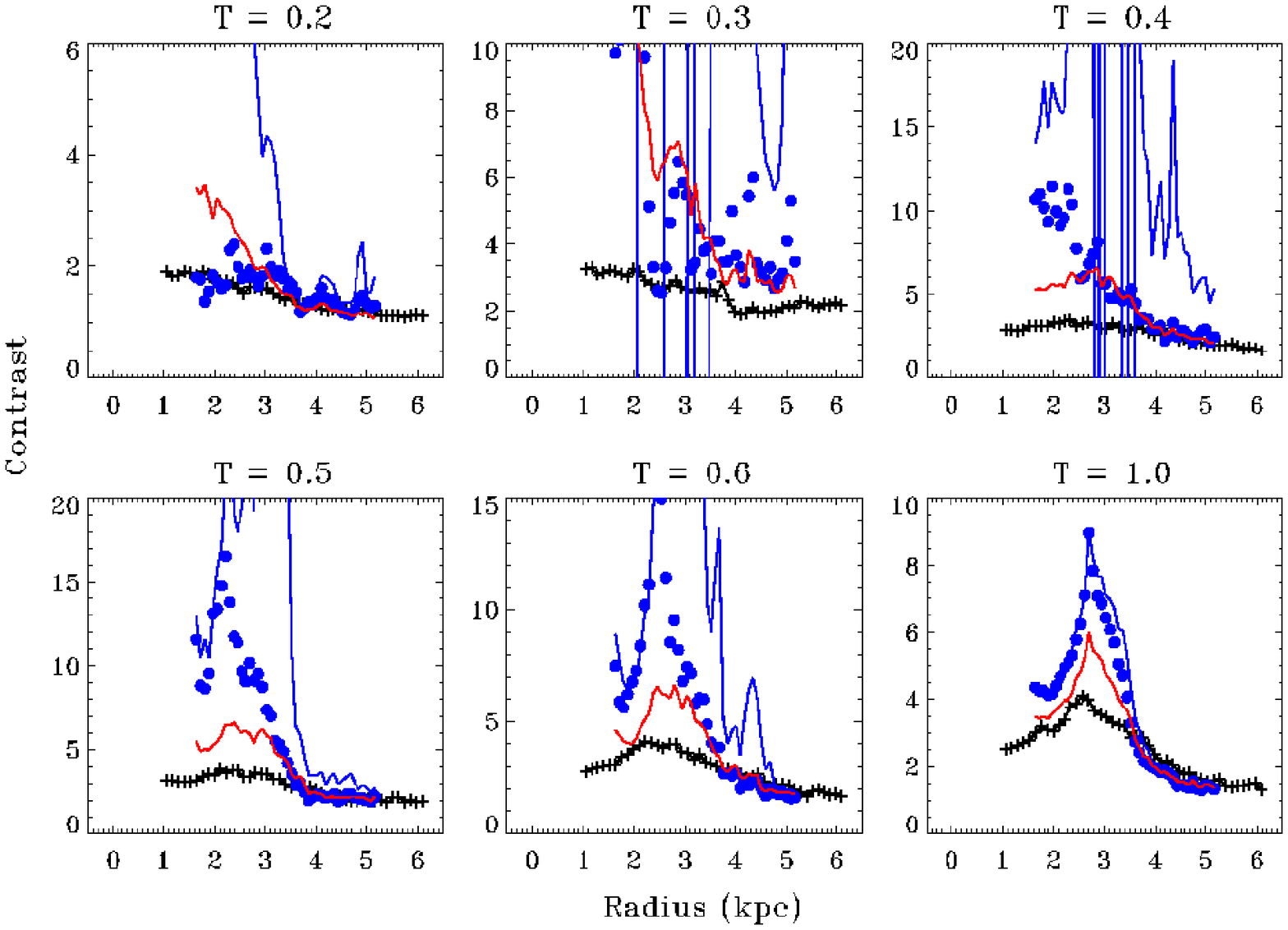}}
\caption{Contrast profiles inside the central 6~kpc of Run \FBDA,
determined on the mass distributions (black crosses), B images with
extinction (blue full circles), dust-free B images (blue line), and K
images (red line). Measurements inside the central kpc are not plotted
since they are not reliable.}
\label{fig:contrast}
\end{figure*}

Each mock image was analysed by fitting simple ellipses to the isophotes.
Profiles of surface brightness, ellipticities, and PA were obtained by
increasing the \smal\ $a$ by a factor 1.01 between each fit. We chose such a
low value to obtain a good resolution in the inner region.  We display the
results for Run \FBDA\ in Figs.~\ref{fig:profiles} and \ref{fig:profiles2} for
the same selected times as in Fig.~\ref{fig:contour}.  Results for the other
runs are qualitatively similar, so they are not displayed.

The ellipticity profiles in both bands are significantly different for $t <
1$~Gyr. In Paper~I, we proposed that, even in the absence of extinction, two
causes are simultaneously responsible for the wavelength dependence of
ellipticity profiles. First, the spatial distribution of the new population is
very elongated along the bar because of new stars born in the gas flow along
the bar that is narrower than the stellar bar. Then, secular evolution is
responsible for progressively making it rounder. This explains the high
ellipticity reached in the bar when the star-formation rate (SFR) is high
($t<0.5$~Gyr for Run \FBDA) and, only in part, the subsequent decrease.
Second, the luminosity ratio between the new and the initial populations is
wavelength dependent, being higher in B than in K band.  After the SFR
maximum, the morphology becomes gradually dominated by the luminosity of the
initial population that has a rounder spatial distribution. When dust
extinction is taken into account, especially in the B band, there is no longer
a unique $e^{\rm max}$ because the real maximum is located in the dustiest
region (e.g. $t=0.3$~Gyr in Fig.~\ref{fig:profiles2}).  A comparison between B
and K-band \cun\ measurements confirms that ellipticity is strongly dependent
on the colour.

\begin{figure}
\centering
\resizebox{\hsize}{!}{\includegraphics{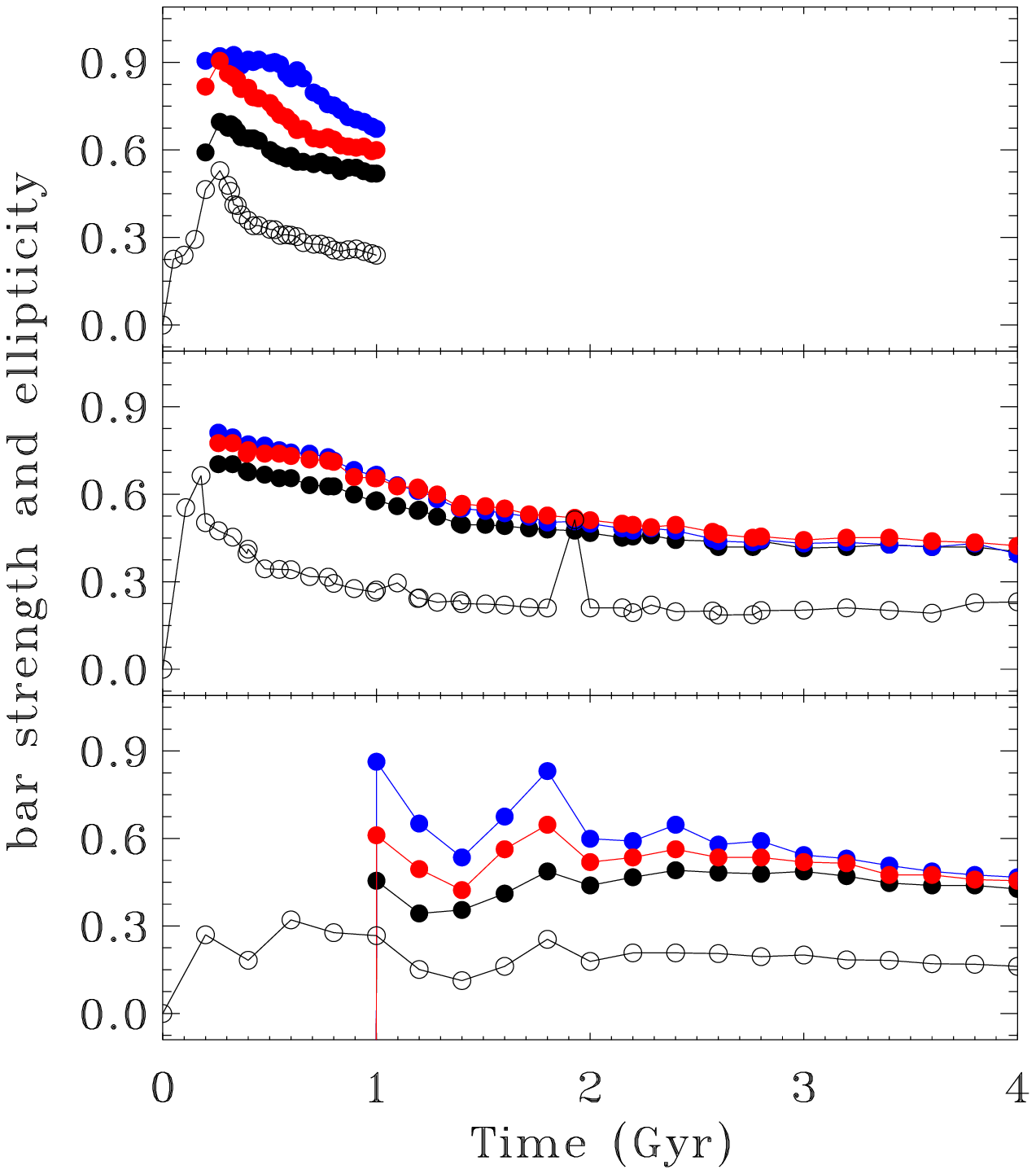}}
\caption{Evolution of the bar strength $Q_b$ (open circles) for Runs
\FBDA\ (top), \FBEA\ (middle), and \WABA\ (bottom). The maximum
ellipticity of the bar is plotted as full circles for the mass density
(black), B images (blue points), and K images (red points).}
\label{fig:barstrengths}
\end{figure}

It is interesting to compare the maximum ellipticity with the bar
strength $Q_b$ defined as \citep{CS81}:
\begin{displaymath} 
Q_b = \max\left( \frac{F_\theta^\mathrm{max}(r)}{<F_R(r)>} \right)
\end{displaymath} 
where $F_\theta^{max}(r)$ is the maximum tangential force at radius $r$ and
$<\!F_R(r)\!>$ is the average radial force from the axisymmetric component.
Recently, this bar strength estimator has been also used in photometric
studies (\citealt{BB01}, \citealt{Betal05}). Comparison of the evolution of
the bar ellipticity with that of $Q_b$ (cf.  Fig.~\ref{fig:barstrengths})
clearly confirms that the bar axis ratio (or ellipticity) is not an estimator
of the bar strength $Q_b$. Indeed, bars with larger quadrupole moments and
lower axis ratios can have the same $Q_b$ as bars with smaller quadrupole
moments and higher axis ratios.  Moreover, the radius at which $Q(r)$ reaches
its maximum, used to define $Q_b$, is always located in the circumnuclear
region.  Thus, $Q(r)$ or $Q_b$ is not useful for determining the bar radius.
The distributions of $Q_b$ for our simulations and for real galaxies
\citep{Betal05} are very similar, except for $Q_b < 0.15$ (very weak bars).
Bar morphology and properties, such as strength and ellipticity, produced in
our simulations are thus representative of real galaxies in the local
universe.

Due to the definition of \cdeux\ and \ctrois, PA variations are $\la 10\degr$
within the radial range defined by \cdeux\ and \ctrois\ criteria. This region
of strong isophotal twist clearly marks the boundary of the bulk of the bar.
Outside this transition region the surface photometry is dominated by the disc
contribution and possibly by spiral arms; however, the size of this twisted
region could be very time dependent, especially in the presence of spiral
arms. This leads to variation from one snapshot to the next in \cdeux\ and
\ctrois\ bar radii since the region could shrink or expand on a short
timescale.

Fourier analysis was also done for each mock image.  Radial profiles of the
contrast computed with B and K images, as well as with the mass distribution,
are displayed in Fig.~\ref{fig:contrast} for Run \FBDA. In general, the
contrast follows a well-defined curve, with a clear maximum once a bar is
formed. However, during the most active star-formation phase, large
fluctuations of the contrast, especially in the dust-free B band, make it
difficult to draw any radial profile. The contrast is also very sensitive to
the extinction in the B band, as opposed to the K band. Interestingly, the
contrast is always higher in the B band than in the K band. The lowest values
are reached for the mass distribution. With another observational tool, this
confirms the result found in Paper~I that the bar shape (ellipticity or
contrast) depends both on the wavelength and on the delay after the major
star-formation phase.

However, it could be difficult or even impossible to define a maximum (e.g.
$t=0.3$ or $t=0.4$~Gyr in Fig.~\ref{fig:contrast}). Moreover, the minimum
could sometimes be negative in the case of very elongated and thin-light
distribution, which is the case when star formation is active along the gas
streamlines in the bar. Indeed, $I_\mathrm{ib}$ takes negative values. This
also explains why we failed to apply the \cquatre\ criterion to dust-free B
images. Sometimes, the contrast remains lower than 2 everywhere (as for the
test image in Sect.~\ref{ssec:fourier}), which makes it impossible to obtain
\cquatre.

\subsection{Projection and resolution effects}
\label{ssec:projection}

Because real galaxies are never exactly face-on, it can be asked how the bar
radius determined is affected by projection effects when using the six
criteria.  We used one snapshot (Run A at $t=0.77$~Gyr) to study, on one hand,
the effects of spatial resolution and, on the other, the effects of
inclination. We used images in K band.

When working with inclined galaxies, we also had to take the PA of the bar 
with respect to the line of node into account. 
Therefore we measured the radius of the bar using the six criteria for
simulated galaxy images inclined from $15\degr$ (nearly face-on) to
$70\degr$. For each inclination, we used 4 values for the bar position-angle
(PA~$=0\degr$, $30\degr$, $60\degr$, and $90\degr$ ).

For $i=15\degr$, the measurements of bar radii are rather insensitive to the
PA variation.  For $i>15\degr$ there are two regimes. For PA~$< 60\degr$, bar
radii decrease with PA, and this decrease is stronger for higher values of
inclination.  For PA~$>60\degr$, the trend depends on the criterion. For
example, bar radii determined with \cquatre\ and \ccinq\ decrease with PA,
whereas they increase for \cdeux. With the \cun\ criterion, the bar radius is
almost constant in this range of $i$ and PA. For \ctrois\ and \csix, bar
radii depend on the degree of inclination. In addition for $i=70\degr$
\cquatre\ does not work, and for high values of $i$ and low values of PA,
\cun\ gives a bar radius close to the values obtained with \cdeux, \ccinq, and
\csix.

To give a rough estimate of the error made on the bar length using imperfectly
deprojected images of inclined galaxies, we find that in all cases the
difference for $i<30\degr$ is less than $20\%$ and the difference for
$i<15\degr$ is less than $10\%$. For our images, with a bar radius at 5 kpc,
$10\%$ represents less than 5 pixels.

Finally one could also wonder how the spatial resolution of the images affects
the determination of the bar length.  We then measured the bar radius using
those six criteria in K-band images with lower resolution (two to five times
lower resolution; the lowest resolution is then approximatively $0.6$~kpc per
pixel).  For a two-times lower resolution, there is no difference: variation
lower than $2\%$, which is lower than incertitudes due to human measurement.
For lower resolution, bar-radius measurements with \cdeux, \cquatre, and
\ccinq\ have small variations ($\approx 5\%$), whereas with \cun, \ctrois, and
\csix\ they vary more significantly ($10 - 15\%$).  When resolution is
decreased, the trends are that bar radius determined with \cdeux\ decreases
slightly, increases slightly with \cquatre\ and \ccinq, increases with
\ctrois, and decreases with \csix. The \cun\ criterion is more fluctuating,
and bar radius is then smaller or longer, depending on the resolution. Due to
these effects, \cun, \cdeux, and \csix\ criteria could give close values for
the bar radius at low resolution.

\subsection{Dust extinction effects}
\label{ssec:extinction}

Extinction effects in the bar region are important. For Run \FBDA\ at
$t\approx 0.3$~Gyr, less than 20\%\ of the bar B-band luminosity escapes,
increasing to 55\%\ at $t\approx 1$~Gyr.  As shown in Paper~I, B isophotes are
more disturbed by dust extinction than K ones. Since our dust model is
proportional to \ion{H}{i} column densities, extinction decreases with time
following the consumption of gas by star formation.  A comparison of
Figs.~\ref{fig:profiles} and \ref{fig:profiles2} shows that ellipticity
profiles are completely different and very noisy when extinction is high,
making bar-radius determination difficult.  Differences in B-band ellipticity
profile are still present at $t=1$~Gyr but mainly appear in the central kpc.

Regions with high SFR are located where gas density is the highest. In these
regions, extinction is then very high since it is proportional to the gas
column density. The regions that contain the youngest and bluest populations
are also the most obscured.  Since the gas is consumed during the evolution
and since gas and stars have different kinematics, the gas density decreases
secularly, enabling the escape of the blue light after several tens of Myr.
However, the gas consumption occurs inhomogeneously, and at various rates, in
the bar region.  

All criteria are not affected by extinction in the same way.  For instance,
the \ctrois\ criterion is in general rather insensitive to the extinction
because the measurements are made in regions where the gas density is lower
than in the innermost regions but, depending on the detailed spatial
distribution of the gas, it could also lead to large errors. For Run \FBDA,
the \ctrois\ bar radius at $t=0.6$~Gyr differs by 1~kpc, whether extinction is
present or not.  It is also noteworthy that, although the \cun\ bar radii
differ when extinction is added to the images, maximum ellipticity values do
not significantly differ from the dust-free case.  For Run \FBEA, the
ellipticity and PA profiles in B band are strongly affected by dust extinction
during the 2 first Gyr. For most of these snapshots, we were not able to
determine the bar radius using \cun, \cdeux, and \ctrois.

Paradoxically, the contrast profiles ${\cal C}(r)$ are much more reliable when
extinction is taken into account, since the very active regions along the gas
streamlines are hidden. Fluctuations of $I_\mathrm{ib}$, the denominator in
the definition of ${\cal C}$, are thus strongly smoothed.  The application of
\csix\ to dusty images leads to much more difficult measurements of bar radii,
in particular in the B band and during the first hundred Myr of evolution when
the gas density is still high.

%-------------------------------------------------------------------
% EVOLUTION DES LONGUEURS DE BARRE
%-------------------------------------------------------------------
\section{Evolution of bar radius}
\label{sec:results}

\begin{figure*}
\resizebox{\hsize}{!}{\includegraphics{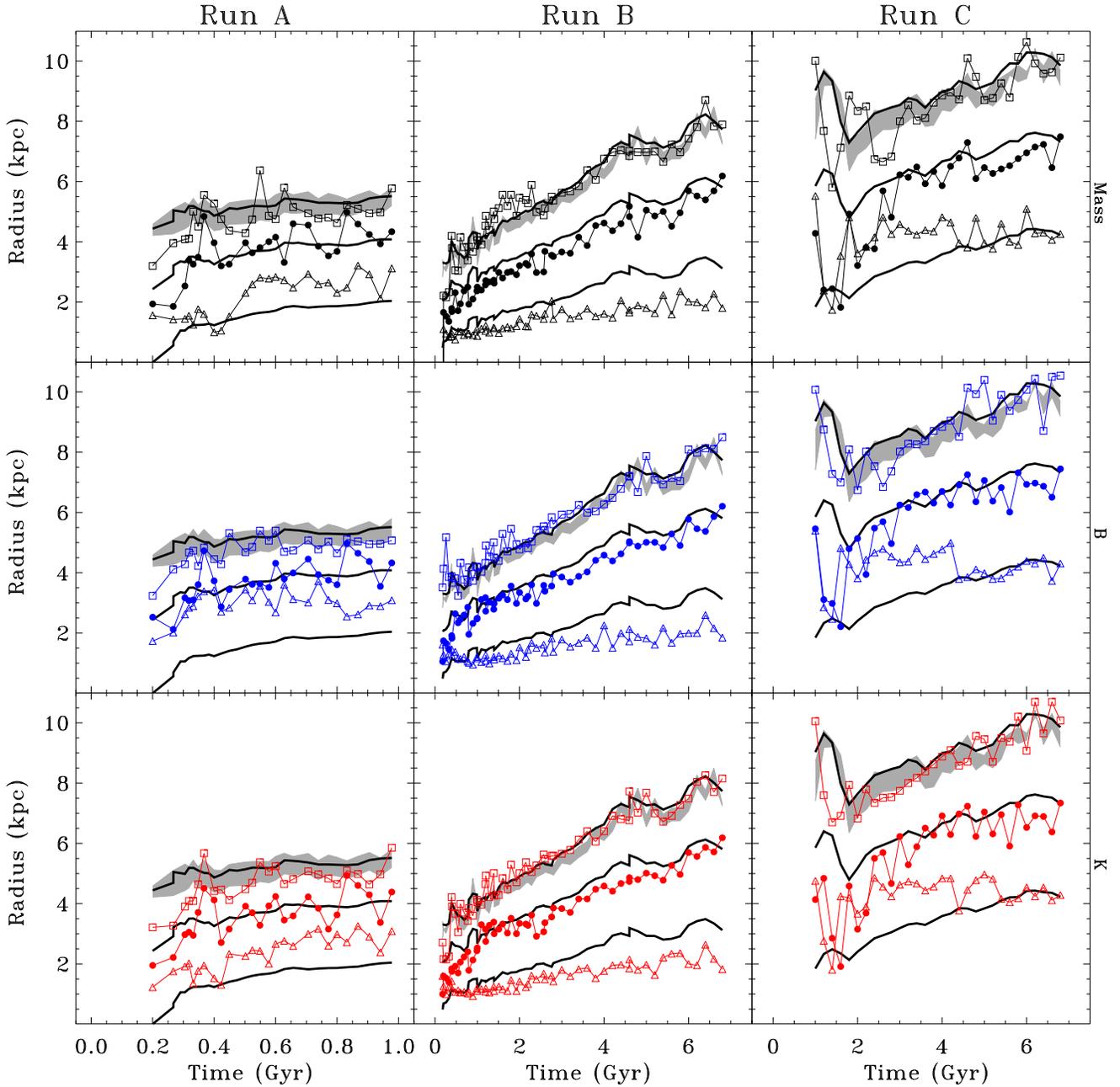}}
\caption{Bar and resonance radii for Runs \FBDA, \FBEA, and \WABA. For
  each panel, lines represent the ILR (shorter radius), UHR (intermediate
  radius), and CR (greater radius).  The dimmed region shows the
  \rlun--\rlquatre\ range. Bar radii determined with the maximum ellipticity
  criterion (\cun) are plotted as open triangles, those determined with the
  minimum ellipticity criterion (\ctrois) are plotted with open squares, and
  the new criterion (\cdeux) is represented by full circles. {\bf Top panels:}
  bar radii determined for the mass. {\bf Middle panels:} B dust-free images.
  {\bf Bottom panels:} K dust-free images.}
\label{fig:barlengths1}
\label{fig:barlengths11}
\label{fig:barlengths12}
\end{figure*}

\begin{figure*}
\resizebox{\hsize}{!}{\includegraphics{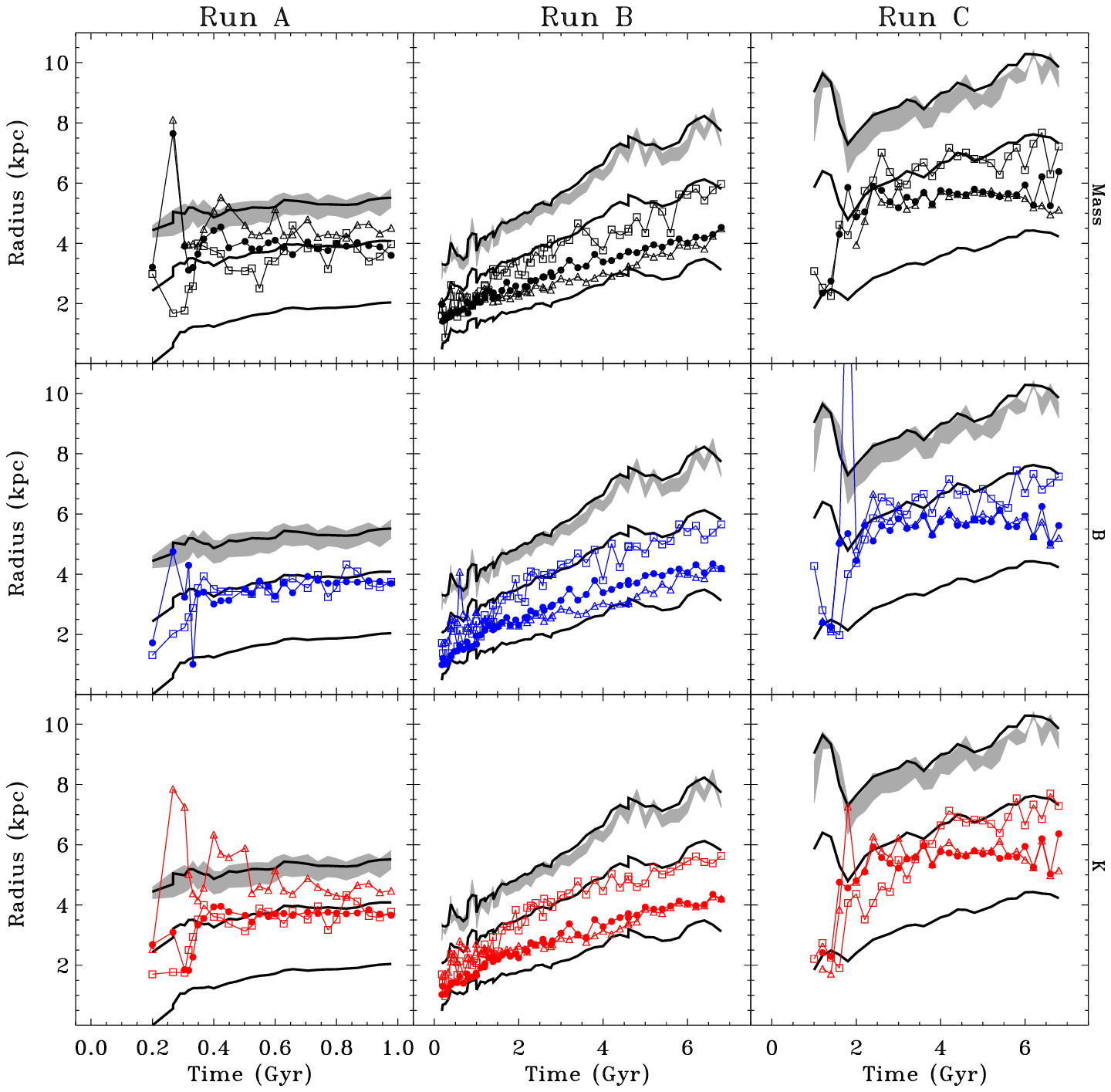}}
\caption{Like Fig.~\ref{fig:barlengths1} but for bar radii determined
  with the Ohta et al.  criterion (\cquatre, open triangles), Aguerri et al.
  criterion (\ccinq, full circles), and $m=2$ phase criterion (\csix, open
  squares). {\bf Top left panel:} measurements for the mass. {\bf Middle
    panels:} B dust-free images. Measurements for $t < 0.4$ Gyr are not
  reliable for Run \FBDA\ (see text).  {\bf Bottom panels:} K dust-free
  images.}
\label{fig:barlengths2}
\label{fig:barlengths21}
\label{fig:barlengths22}
\end{figure*}

The evolution of the bar radius, determined using all six criteria, is plotted
in Figs.~\ref{fig:barlengths1} and \ref{fig:barlengths22}. In the case of Runs
\FBDA\ and \FBEA, for $t \la 0.2$~Gyr, the bar progressively grows but is
still too weak to give a sharp signature in the surface-brightness
distribution. Thus, each of the criteria fails to accurately define a bar
radius during this growing phase.  This is also the case for Run \WABA\ for $t
\la 1$~Gyr, since the bar formation has a longer timescale due to the more
massive initial bulge.

We hereafter discuss the results obtained for Run \FBDA\ and then look into
details of the evolution on the longer term.

\subsection{Results with \cun, \cdeux, and \ctrois}

Criteria based on ellipse fitting give different estimations of the bar
radius; \cun\ gives generally lower values, while \ctrois\ gives the highest
estimation, as expected (cf. Sect.~\ref{ssec:ellipse}). However, \cun\ and
\cdeux\ in B band give similar results during the active star-formation phase
($t \la 0.6$~Gyr) because of the steep ellipticity gradient due to the sharp
bar boundary.

The bar radii for the mass density and K images show the same trend as a
function of time. They thus could be considered as similar on average,
although results clearly show larger fluctuations for the mass density than
for the K images.

For the three criteria, the large fluctuations of bar radius from one snapshot
to the next do not have a dynamical origin. This is only the consequence of
our attempt to apply the criteria in the most objective way. Indeed, we
systematically disregard bar-radius determination for previous snapshots.
These fluctuations must be considered as an estimation of the error caused by
using a criterion based on ellipse fitting. Since an ellipticity maximum is
easier to determine than any other peculiar place on ellipticity or PA
profiles, \cun\ gives less noisy estimations than \cdeux\ or \ctrois.

\subsection{Results with \cquatre, \ccinq, and \csix}

In comparison with the previous criteria, \cquatre, \ccinq, and \csix\ give
different radius estimations, evolution, and level of fluctuations.  Indeed,
the bar radius determined with \cquatre\ and \ccinq\ is roughly constant with
a little dispersion for $t > 0.5$~Gyr, whereas it increases slightly with time
and with a stronger dispersion for the other ellipse-based criteria. Bar
radius determined with \csix\ first increases, then slightly decreases for $t
> 0.8$~Gyr. For \cquatre\ and \ccinq, the level of fluctuation from one
snapshot to the next is lower than in the case of \cdeux\ and \ctrois\ 
because: 1) these criteria do not rely on human decision and 2) high
frequencies in the image are smoothed out. However, \csix\ is a bit noisier
than \cquatre\ or \ccinq\ probably because this criterion also relies on human
decision as do the \cdeux\ and \ctrois\ criteria.  Moreover, \cquatre\ and
\ccinq\ give similar results for the mass but differ on K-band images,
\cquatre\ giving a longer bar than \ccinq\ by $\approx 1$~kpc.

The \ccinq\ criterion deserves particular attention since it gives similar and
reliable results for the mass, B, and K images for $t > 0.5$~Gyr, that is,
0.2~Gyr after the peak of SFR. For $t < 0.3$~Gyr, when the bar is still
growing, bar-radius estimation is unreliable. For $0.3 < t < 0.5$~Gyr, when
star formation is the most active, the bar radius is shorter in B than in K,
the highest value being obtained for the mass. However, the contrast ${\cal
  C}$ profiles (Fig.~\ref{fig:contrast}) clearly show that these estimations
are not reliable, apart from the mass density.

Let us recall that it was not possible to get valuable bar radii with
\cquatre\ on B-band images because the contrast often remains above the
threshold (fixed at 2 by Ohta et al.) until the end of the bar.

\subsection{Long-term evolution}

All the above analysis were applied to the other two runs that were performed
until $t=7$~Gyr. We reached roughly similar conclusions when we compared the
results of the three runs. Such a comparison does make sense only on condition
that the different evolution histories of each run are taken into account. The
first Gyr of Run \FBDA\ is therefore comparable to almost the first 1.5 Gyr of
Run \FBEA\ and the first 3 Gyr of Run \WABA.

The differences between Run \FBDA\ and the two other runs mainly concern the
criteria \cun\ and \cquatre\ to \csix.  For Run \FBEA, the \cquatre\ criterion
can be measured from B-band images, and the results are similar to those in K
band. For Run \WABA, the measurements using this criterion are possible and
reliable for $t \ga 2$ Gyr and give the same values as the \ccinq\ criterion.
Another difference concerns the \cun\ criterion for Run \WABA. Indeed, it
gives approximatively the same values as the \cdeux\ criterion for $t \la 3$
Gyr. The \csix\ criterion gives a significantly greater bar radius than those
obtained with \cquatre\ and \ccinq. This criterion is very comparable to
\cdeux.

On the other hand, for the criteria \cdeux, \ctrois, and \cquatre\ to \csix,
each gives similar results when we determine bar radius on the B or K-band
images or on mass distributions, taking the uncertainties of measurements into
account.

%--------------------------------------------------------------------
% DISCUSSION
%-------------------------------------------------------------------
\section{Discussion}
\label{sec:discussion}

\subsection{Comparison with dynamical radii}
\label{ssec:compare}

In Fig.~\ref{fig:barlengths1} we also plot the radius of the CR (\rcor), ILR,
and UHR resonances.  In order to estimate the dynamical resonances radii, we
computed the circular orbit frequency $\Omega$ and the radial epicyclic
frequency $\kappa$ as \citep{P90}:
\begin{displaymath}
\Omega^2(r) = \left\langle
\frac{1}{r}\frac{\partial\Phi}{\partial r}
\right\rangle
\end{displaymath}
\begin{displaymath}
\kappa^2(r) = \left\langle
\frac{\partial^2\Phi}{\partial x^2}+
\frac{\partial^2\Phi}{\partial y^2}+
2\,\left( \frac{1}{r}\frac{\partial\Phi}{\partial r} \right)^2
\right\rangle
\end{displaymath}
where $\Phi$ is the gravitational potential and $\langle\cdots\rangle$
stands for an azimuthal average.

Strictly speaking, these frequencies predict the oscillation frequencies of
the orbits only in the axisymmetric case. They do not indicate whether
families of periodic orbits actually follow such oscillations when the bar
growth breaks the axisymmetry. In fact, a proper orbital analysis is more
appropriate. However, a number of previous orbital studies incline us to
consider that the epicyclic approximation could lead to an acceptable
estimation of the resonance locations.

Indeed, for the ILRs, \citet{A92a} has computed the spatial extension of the
$x_2$ and $x_3$ families associated with the existence of these resonances.
Her comparison with ILR radii calculated by averaging the frequencies over the
azimuthal angle shows that the error cannot exceed 10\%\ in the worst cases.
Since our $N$-body models are well inside the parameter space used by
\citet{A92a}, we confidently use the linear approximation to estimate the
radius of the outer ILR.

To our knowledge, the differences between the real Lagrangian points location
and the axisymmetric corotation radius have been less quantified in the past.
In the case of barred galaxies, Lagrangian points are defined as the points of
equilibrium between centrifugal and centripetal forces along the main axis of
the bar perturbation. The radii of these points converge towards the
corotation circle when the bar perturbation vanishes. Following \citet{P90}
and assuming a perfect bisymmetric barred potential, we call \rlun\ and
\rlquatre\ the radii of the Lagrangian points along the major-axis and
intermediate axis (minor-axis in a face-on projection), respectively, of the
bar perturbation. In a strong bar potential, \rlquatre\ is slightly smaller
than \rlun. However, the amplitude of the difference between \rlun\ and
\rlquatre\ should be roughly proportional to the bar strength.

\begin{figure}
\centering
\resizebox{\hsize}{!}{\includegraphics{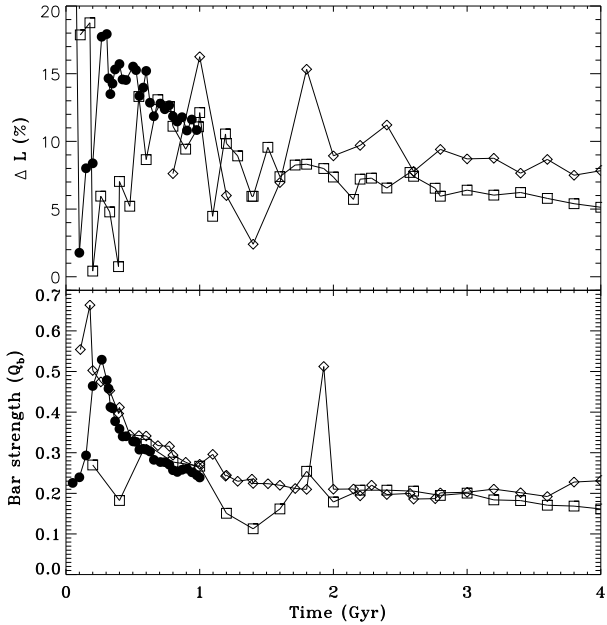}}
\caption{Evolution of the relative difference between \rlun\ and
\rlquatre\ (top) and bar strength $Q_b$ (bottom) for Runs \FBDA\ (full
circles), \FBEA\ (lozenges), and \WABA\ (squares).}
\label{fig:rl}
\end{figure}

In Fig.~\ref{fig:rl} we thus display the quantity $\Delta L =
({L_{1,2}-L_{4,5}})/\overline{L}$, where $\overline{L}$ is the average between
\rlun\ and \rlquatre.  The relative amplitude of the difference between \rlun\ 
and \rlquatre\ rarely exceeds 15\%\ even in very strong bar phases, as for
instance during the bar-formation phase in Run \FBEA. A standard value for a
slowly evolving bar seems to be in the range 5 to 10\%. A comparison with the
$Q_b$ evolution clearly confirms this expected trend. When the bar gets
stronger, the difference between \rlun\ and \rlquatre\ increases.

In the epicyclic approximation, corotation is the radius at which
$\Omega(R_\mathrm{CR}) = \Omega_b$, where $\Omega$ is the circular
orbit frequency and $\Omega_b$ is the bar pattern speed. The value of
$\Omega_b$ was accurately determined using the Tremaine-Weinberg method
\citep{TW84} since it has been shown that the method is very reliable on
stellar velocity fields \citep{HWetal05}, i.e. the same values than those
obtained using the phase of the bar. In Fig.~\ref{fig:barlengths1} we compare
the corotation radius with \rlun\ and \rlquatre. As expected, the corotation
radius lies between the Lagrangian points but is closer to \rlun\ than
\rlquatre. Thus, we can use the linear approximation to compute the corotation
radius and $\Delta L$, as defined above, could be used as an estimator of the
error made on the corotation location.

Keeping in mind a typical error of 10\%, as discussed above, we have
determined the position of the radial ILR defined as the solution of
$\Omega(R_\mathrm{ILR}) - \kappa(R_\mathrm{ILR})/2=\Omega_b$ and UHR for which
$\Omega(R_\mathrm{UHR}) - \kappa(R_\mathrm{UHR})/4=\Omega_b$.

The last check was to compute the orbital frequencies $\kappa$ and $\Omega$ of
a representative sample of particles. We applied a variant of the technique of
\citet{A03}. We froze the potential at a given time in the simulation and then
computed orbits in the inertial potential to determine the main frequencies
using the technique of \citet{CA98}. We used between 100\,000 and 200\,000
particles as initial conditions chosen at random. This leads to the
determination of $\kappa$ and $\Omega$ for each particle which can be plotted
in a $\kappa-\Omega$ diagram. Resonant families (e.g. ILR) and corotation are
thus easily identifiable. We repeated this very CPU time-consuming
computation\footnote{It takes roughly two weeks by snapshot on a 12-node
  cluster for 200\,000 orbits} for 4 snapshots in order to check the
consistency with the linear approximation described above. We thus determined
that the error on the corotation radius never exceeds 10\%.

The resonance locations are obviously independent of our photometric
calibration and thus could be compared to our estimation of bar radii using
various criteria.  However, for all the reasons mentioned in
Sect.~\ref{sec:tests} about $e^\mathrm{max}$, \cun\ cannot be, a priori, a
good estimator of any dynamical radius (ILR, CR, etc.). This is clearly
confirmed in Fig.~\ref{fig:barlengths1}.  Also, \ctrois\ gives bar radii that
are very close to the corotation. This strongly suggests that the ellipticity
minimum at the end of the bar could be due to the signature of the corotation
radius. However, if we define the corotation region as between \rlun\ and
\rlquatre, \ctrois\ gives a lower limit of the corotation.

The criteria \cdeux, \ccinq, and \csix\ give bar radii that are rather well
correlated with the UHR radius. However, \ccinq\ is less correlated when SFR
is high (for $t < 0.5$~Gyr), whereas \cdeux\ could give a higher error on the
UHR position. One more difference appears between various simulations for the
\ccinq\ criterion. For Run \WABA, it is correlated with the UHR radius for $t
\la 3.5$ Gyr, i.e. for the duration comparable to Run \FBDA. On longer
timescales, this is no longer the case. For Run \FBEA, the \ccinq\ criterion
never appears to be correlated with the UHR radius, and it gives values
between the ILR and the UHR radius.  Nevertheless, \cdeux\ and \csix\ criteria
are still correlated with the UHR radius on longer timescales, like the
\ctrois\ criterion with the corotation radius.

The fact that \cdeux, \csix, and in some cases \ccinq\ criteria are good
estimators of the UHR radius during the first Gyr can be understood in the
framework of stellar orbits. Indeed, it has been shown \citep{C81a} that the
UHR (or 4/1 resonance) is a gap where no family of periodic orbits can exist.
Moreover, between the UHR and the corotation, most families of periodic orbits
are unstable \citep{C81b} if the bar perturbation is stronger than 10\%\ of
the potential background (the stellar disc). Using 3D models, \citet{SPA02}
also conclude that the most appropriate orbits to sustain a bar are those
inside the UHR.  Thus, between the UHR and the corotation, a bar is mostly
sustained by semi-chaotic orbits \citep{WP99}. The density response of these
semi-chaotic orbits in the configuration space is rounder than the response
density of orbits trapped around the major families of periodic orbits (i.e.
$x_1$ and 3/1 resonant family). Thus the contrast as defined by
Eq.~(\ref{eq:contrast}) should strongly decrease after the UHR. This also
explains why \ctrois\ is a good tracer of the corotation radius.

However, for Run \WABA, the \ccinq\ criterion is no longer correlated with the
UHR radius from $t \ga 3.5$~Gyr and never appears to be correlated with the
UHR radius for Run \FBEA. This means that this criterion is not a reliable
estimator of the resonance location. The bar--interbar contrast indeed shows
the pattern of the bar for all the snapshots of these three runs, but the
decrease of the contrast at the end of the bar, which is the transition region
between the bar and the disc, seems to be very sensitive to local conditions
(such as star formation at the end of the bar, the spiral structure in the
disc, and the thickness of the disc); therefore, these measurements depend on
the type of bar and evolution.  Then, even if the \ccinq\ criterion is an
automatic and reliable criterion of the bar radius, it defines a radius that
is not always correlated with a resonance and then is not comparable from one
to another galaxy.

\subsection{Fast and slow bars}
\label{ssec:fastslow}

\begin{figure}
\resizebox{\hsize}{!}{\includegraphics{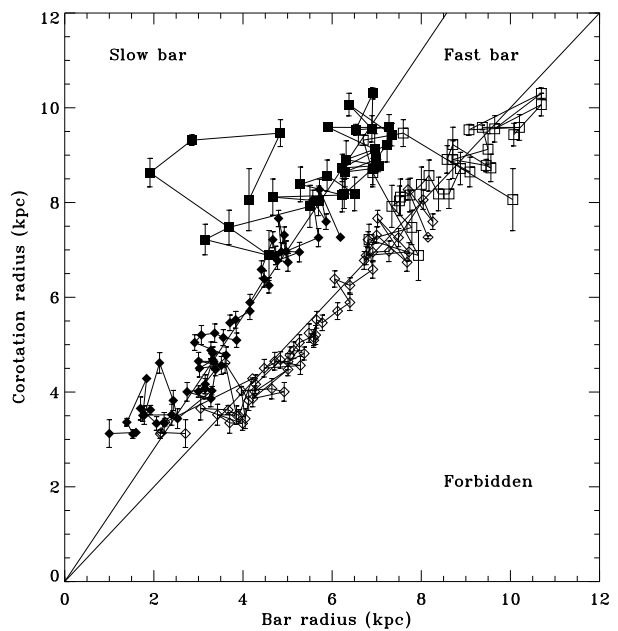}}
\caption{Plot of corotation radius versus bar radius for Runs \FBEA\ (lozenges)
  and \WABA\ (squares). The corotation radius was estimated using
  $\overline{L}$, and the error is given by the difference between \rlun\ and
  \rlquatre. For the bar radius, we used criteria \cdeux\ (full symbols)
  and \ctrois\ (open symbols), determined on K-band images.}
\label{fig:fastbar}
\end{figure}

We plotted the corotation versus the two more accurate determinations of bar
radius (cf. Fig.~\ref{fig:fastbar}).  The corotation radius was estimated
using $\overline{L}= (L_{1,2}+L_{4,5})/2$ as in Sect.~\ref{ssec:compare}. The
error is given by the difference between \rlun\ and \rlquatre.

Using such a plot, \citet{Aetal03} have divided fast bars 
($1 < {\cal R}=R_{\mathrm{CR}}/R_{\mathrm{bar}} < 1.4 $) 
from slow ones (${\cal R} > 1.4$). For our simulations, especially for the
long Runs \FBEA\ and \WABA, ${\cal R}$ does not evolve very much with time
except during the bar-formation phase.  Once the bar is settled in, ${\cal R}$
remains rather constant. 

It is noteworthy that the value of ${\cal R}$ is very sensitive to the
definition of the bar radius. For the same simulation (either Run
\FBEA\ or \WABA), \cdeux\ (or \csix) leads to the conclusion that the bar is a
rather slow rotator, while \ctrois\ leads to the opposite conclusion. There is
unfortunately no obvious theoretical definition of the `radius' or `length' of
the bar. The only clean theoretical prediction is that bars cannot extend
outside the corotation radius \citep{C80}, whereas they could be limited by
the disc scale length and thus could end near the ILR \citep{CE93}.
Moreover, when the ratio of the bar radius to the corotation radius is given
(e.g. $1.2\pm 0.2$, cf. \citealt{A92}), the bar radius is in fact defined as 
the half-major radius of a Ferrers ellipsoid that is not an observable radius,
so that any comparison with observational bar radii could be difficult.

In the case of the measurement of ${\cal R}$, the use of \ctrois\
should obviously be avoided. Indeed, \ctrois\ points out the
corotation, leading to a ratio ${\cal R}\approx 1$. Thus, \cdeux\ and
\csix\ criteria, being loosely correlated to a dynamical radius (UHR),
could be useful in a first estimation of a bar radius. Defining
the bar radius in such a way could be considered as a loose definition
of the bar radius since the UHR does not limit the extension of the
bar. We can thus imagine converting $R_\mathrm{UHR}$ into
$R_\mathrm{CR}$, but the ratio $R_\mathrm{CR}/R_\mathrm{UHR}$
decreases from 1.5--1.8 during the short bar-formation phase to
$\approx 1.3$ when the bar is settled in. Thus, the location of the
corotation could only be retrieved within an error of the order of
10~\%. Of course, the combination of the \cdeux\ criterion with \ctrois\
could lead to a better estimation of the corotation radius.

\subsection{Dark-halo effects}

The main effect of a live dark halo (except to flatten the rotation curve of
the disc at large distance) is to permit exchanges of angular momentum with
the stellar disc. The rate and amount of these exchanges depend on the
velocity dispersion of both the disc and the halo, and on the relative halo
mass (cf. \citealt{DS00}, \citealt{A03}, \citealt{VK03}). The stellar disc
could lose between a few \%\ and 40\%\ of its angular momentum. Depending on
the rate at which the stellar disc loses its angular momentum, the bar grows
quite differently. The lack of dark halo in our simulations thus has mainly
three consequences on our work:

1) The evolution timescale could be different from simulations with a dark
halo, leading us not to draw any conclusion about the evolution of ${\cal R}$,
for instance.

2) The shape and strength of a bar are also affected by the presence of the
live halo \citep{AM02}. The central concentration of the halo is thus a key
parameter. For instance, \citet{AM02} stress that more centrally concentrated
halos lead to rectangular-like bars.  However, the photometric and
morphological properties (shape, ellipticity, $Q_b$, etc.; cf. also Paper~I)
of our simulated bars are very representative of real stellar bars. For
instance, we are able to reproduce a rectangular-like bar (Run \FBEA,
Fig.~\ref{fig:contour2}).

3) The main results that could be affected by the absence of a dark halo in
our simulations are those related to dynamical properties, i.e.  the
correlation of the position of resonances with bar-radii
estimators (Sect.~\ref{ssec:compare}).  This is a major concern since there is
a debate on whether a dark halo increases ${\cal R}$ or not, leading to
significantly smaller bars compared to corotation. \citet{DS00} report high
values of ${\cal R}$ for a number of their simulations.  However, \citet{DS00}
estimate bar radii using in part the \csix\ criterion that is, for our
simulations, quite correlated to UHR. This criterion gives rather short bars,
since it is mainly influenced by the part of the bar sustained by $x_1$
orbits.  This might explain the high ${\cal R}$ values in part. For the
simulations by \citet{OD03} and \citet{VK03}, ${\cal R}$ remains in the range
1.2$-$1.7. \citet{VK03} mainly use a criterion based on surface density (not
tested in the present paper), which seems to give similar results than \csix,
whereas \citet{OD03} use an average of three methods (including \cdeux\ and
\csix).

Our results should thus be confirmed with simulations that include a live dark
halo. However, the presence of a dissipative component that can also
exchange angular momentum and mass with the collisionless components through
gravitational interaction, star formation, and feedback processes must have
some influence on the global angular momentum exchanges. Therefore, realistic
models should now include both dark matter and a dissipative component. This
will be published elsewhere.

%--------------------------------------------------------------------
% CONCLUSION
%--------------------------------------------------------------------
\section{Conclusions}

We have used N-body simulations, including stars, gas, and star
formation, that were photometrically calibrated in B and K wavebands to compare
the bar lengths determined by six observational criteria (five
commonly applied to observations or simulations and a sixth
introduced in this paper). These criteria are based on ellipse fitting
(\cun\ to \ctrois) and Fourier analysis (\cquatre\ to \csix) of the
surface brightness (or mass surface density).  The bar-length estimates
were also compared to the location of resonances.

We have obtained the following main results for each criterion in turn:
\begin{itemize}
\item \cun\ \citep{WP91}, the radius of the maximum ellipticity, gives
the shortest bar lengths. It clearly underestimates the real bar
radius. Moreover, it is not linked to any dynamical characteristics of
a bar (e.g. resonance).
\item \cdeux\ is measured at the end of the PA plateau.
It can be used to approximately locate the UHR. 
\item \ctrois, the radius of the minimum of ellipticity
\citep{Wetal95}, gives the greatest bar lengths. In general, it is less
sensitive to dust absorption than other criteria. This is the only
criterion, amongst the six criteria tested, that is correlated with
the corotation.
\item \cquatre\ \citep{o90} is not always defined, since the
bar--interbar brightness contrast ($\cal{C}$) could remain under 2 even
for strong bars.
\item \ccinq\ \citep{Aetal00}, also based on $\cal{C}$, is the least
noisy estimator. However it gives contradictory results for our
simulations.
\item \csix, widely used to determine the bar length in N$-$body
simulations (e.g. \citealt{DS00}), gives bar lengths comparable to
\cdeux\ results. Like \cdeux, it gives quite a good approximation of the
UHR location.
\end{itemize}

In general, for projection angles $i$ between 15 and 70\degr, the bar
length is underestimated with respect to its face-on value, apart from
using \cun. However, for $i<30\degr$, errors remain below 20\%.
Measurements also depend on the resolution but errors never exceed
15\%\ when resolution is downgraded by a factor~5.

We moreover confirm that ellipticity profiles of B and K-band images
can be very different. They also differ from ellipticity profiles of
the mass density. This leads to a clear dependence of the maximum
ellipticity on the waveband. However the radius of the maximum
determined by \cun\ is less affected. Within a typical error of 10\%,
bar lengths determined with other criteria are not colour dependent,
in the absence of dust.  The situation is different when dust absorption
is taken into account. Criteria giving the smallest bar-length
estimate are much more affected than others. This is the case of \cun\ in
particular. 

For each simulation, the maximum ellipticity decreases with time, as
do the bar strengths $Q_b$, but not at the same rate. The use of the
maximum ellipticity to estimate the bar strength could thus lead to
severe errors, especially in a comparison of one galaxy to another.

Accurate determination of the bar length is crucial in several
observational or theoretical analyses. For instance, the ratio
${\cal R}=R_{\mathrm{CR}}/R_{\mathrm{bar}}$ could increase from 1 (fast
bar) to 1.4 (slow bar) just by using \cdeux\ or \csix\ instead of
\ctrois. However, being correlated to the corotation radius, \ctrois\
cannot be used to define the end of a bar. Using \cdeux\ or \csix\ to
define the end of the bar has physical implications since these
criteria point out the UHR. Indeed, inside the UHR, the bar is mainly
sustained by $x_1$ or other families of orbits elongated along the bar
major-axis. However, there is clear evidence that what is commonly
called a bar extends a bit outside the UHR because the region
between UHR and CR is populated by higher resonant families of orbits,
as well as semi-chaotic and chaotic orbits that contribute to the
shape (sometime rectangular-like) of the bar. 
The definition of the bar length (hence the criterion used) should
thus depend on the application.

%--------------------------------------------------------------------
% ACKNOWLEDGMENTS
%--------------------------------------------------------------------
\begin{acknowledgements}
We are grateful to Luis Aguilar and an anonymous referee for comments
and suggestions that have helped to strengthen the conclusions of this paper.
We would also like to thank Luis Aguilar for providing his code to
compute orbital frequencies. Our computations were partly performed on
the Fujitsu NEC SX-5 hosted by IDRIS/CNRS and the CRAL 18-node
cluster of PC funded by the INSU/CNRS (ATIP \# 2JE014 and Programme
National Galaxie).  LMD acknowledges support from a grant from the Universidad
Nacional Aut\'onoma de M\'exico (UNAM) for part of this work. 
\end{acknowledgements}

\end{document}